\documentclass[amsmath,amssymb, reprint]{revtex4-1}

\usepackage[normalem]{ulem}
\usepackage{amsmath,amssymb,gensymb}
\usepackage{array}
\usepackage{graphicx}
\usepackage{dcolumn}
\usepackage{bm}
\usepackage{xcolor}
\usepackage[utf8]{inputenc}
\usepackage[T1]{fontenc}
\usepackage{mathptmx}
\usepackage{etoolbox}
\usepackage{listings}
\usepackage{lmodern}
\usepackage{mathpazo}
\usepackage{microtype}
\usepackage{booktabs}
\usepackage{bm} 
\usepackage{amsmath}
\usepackage{amsbsy}
\usepackage{tabularx}  

\makeatletter
\def\@email#1#2{%
 \endgroup
 \patchcmd{\titleblock@produce}
  {\frontmatter@RRAPformat}
  {\frontmatter@RRAPformat{\produce@RRAP{*#1\href{mailto:#2}{#2}}}\frontmatter@RRAPformat}
  {}{}
}%
\makeatother
\begin{document}

\preprint{}

\title{{\itshape In silico} design and prediction of metastable quaternary phases in Cu-Ni-Si-Cr alloys}

\author{Ángel Díaz Carral}
\affiliation{Institute for Computational Physics, Universit\"at Stuttgart, Allmandring 3, 70569 Stuttgart, Germany}
\email{adiazcarral@gmail.com}
\author{Simon Gravelle}
\affiliation{Univ. Grenoble Alpes, CNRS, LIPhy, 38000 Grenoble, France}
\author{Maria Fyta}
\affiliation{Computational Biotechnology, RWTH Aachen, Worringerweg, 52074 Aachen, Germany}


\begin{abstract} 
Quaternary phases formed in copper alloys are investigated through a combination of quantum-mechanical and classical computer simulations and active machine learning. Focus is given on nickel, silicon, and chromium impurities in a copper matrix. The analysis of the formation enthalpies of candidate quaternary structures leads to the prediction of two novel quaternary phases and the assessment of  their stability. For the predicted two phases, machine learned atomistic potentials are developed using active learning with a quantum-mechanical accuracy. Use of these potentials in atomistic simulations further elucidates the structure, temperature-dependent dynamics, and elastic behavior of the predicted quaternary phases in copper alloys. The combined {\itshape in silico} approach is thus proven highly efficient in both designing materials and elucidating their properties and potential combining different spatiotemporal scales. In the case of alloys, this computational scheme significantly reduces the effort in searching the huge chemical space of possible phases, enhancing the efficiency in synthesizing novel alloys with pre-defined properties.
\end{abstract}

\maketitle 

\section{\label{sec:intro}Introduction}

Complex phases of impurities in alloys, especially in Cu-based alloys, are of high interest due to their conducting properties, as well as strength  \cite{rossiter_1987}. Typically, different impurities can be found in copper alloys, ranging from silicon and phosphorus up to palladium and silver \cite{butterworth1992survey,li1991oxidation,zhou2016precipitation,benedeti1995electrochemical}. Such alloys can find applications in different areas such as in high-heat-flux \cite{minneci2021copper} and biomedical \cite{konieczny2012antibacterial} applications, as high-strength and high-conductivity materials \cite{li2014development}, in pumps, propellers and bearings \cite{izadinia2012microstructural,liu2016effects}, and many more. The use of copper alloys in these strongly relies on the fact that their properties can be tuned by a proper selection of the impurity types and their stoichiometry in these materials. It is known, for example, that copper alloys including  nickel (Ni), silicon (Si), and chromium (Cr) can enhance the strength of the alloy \cite{LI2020102819,gao2020}.
Still the need for further optimizing copper alloys especially with respect to their strength \cite{yu2023progress}, electrical conductivity  \cite{zhou2023enhanced} or both \cite{chu2023simultaneously} is actively being followed. To this aim, known trends need to be further evaluated and build upon. For example, Cu-Ni-Si-Cr alloys  offer an effective barrier for dislocation motion and strength enhancement in the alloy \cite{precipitates1}. More precisely, in alloy binaries of the form Cu-Si, Ni-Si, Cr-Si, Cu-Ni or Cu-Cr, the occurrence of Cr, Ni and Si form clusters and intermetallic phases inside the Cu matrix, which strongly block the dislocation motion in the material  \cite{etde_20026637, Jin1998AgeingCO, cunisi1, LEI201377, wangcunisi}. On the other hand, dissolved impurities in a matrix decrease the electrical conductivity \cite{decelec1,decelec2,decelec3}.

At the same time, high-entropy compounds found in the ternary Cu-Ni-Si \cite{XIE2009114, Tao_2022} or Ni-Si-Cr \cite{schuster, crnisiliu} complexes play an important role in designing copper-based alloys. During aging at 500 $\degree$C, a Cu-Ni-Si-Cr alloy was examined using high-resolution transmission electron microscopy and scanning transmission electron microscopy, revealing the simultaneous presence of three types of precipitates \cite{CHENG2014189}. These phases were identified as ordered face-centered cubic $\beta$-Ni$_{3}$Si, orthorhombic $\delta$-Ni$_{2}$Si, and an ordered face-centered cubic (Ni, Cr, Si)-rich phase.  The smaller (Ni, Cr, Si)-rich clusters, which also contain a substantial amount of Cu (around 50 \%), exhibit an ordered face-centered cubic (fcc) structure and possess a remarkably small size of around 5 nm. Due to these characteristics, these clusters can be regarded as metastable phases. Over an extended period of aging, they eventually undergo a transformation into the $\delta$-Ni$_{2}$Si phase \cite{CHENG2014189,LEE201462}. During thermal aging, precipitation processes can occur depending on the alloy stoichiometry. These  can reduce the number of dissolved impurities significantly \cite{MAIER2019619}.

Therefore, efforts have been focused on developing copper-based alloys that possess both high electrical conductivity and strength. Specifically, in the field of large-scale production of integrated circuits \cite{wang2019}, there is a need for the development of high-performance copper alloys. A better understanding of intermetallic phases in Cu-Ni-Si-Cr alloys is essential in future attempts for the enhancement of their mechanical and electrical properties. To date, no intermetallic phases have been reported in Cu-Ni-Si-Cr alloys  based on either first principles calculations or experiments \cite{chakrabarti1984cr,tang2022,WAN2021158531}. However, more complex compounds found in the ternary Cu-Ni-Si \cite{xie2009microstructure,Tao_2022} or Ni-Si-Cr \cite{schuster,crnisiliu} complexes play an important role in the design of copper-based alloys for simulations. To this end, experimentally synthesized alloys should be extended by theoretically designed and predicted materials. These theoretical approaches can quickly and cost-effectively screen a large number of material candidates. The numerical prediction and simulation of such quaternary structures is the goal of this article. Our work contributes along this way, based on the methodology described in Section \ref{sec:method}, the discussion of the results and prediction of novel phases in copper alloys in Section \ref{sec:result}.

\begin{table}
\centering
\caption{Chemical formula, cell size, and respective number of samples of the unique stoichiometries for the quaternary structures generated in ENUMLIB for the parent fcc lattice.}
\setlength{\tabcolsep}{6pt}
\begin{tabular}{|p{3.cm}|p{1.4cm}|p{2.8cm}|}
\hline
{chemical formula} & {cell size} & {number of samples} \\
\hline
CuSiNiCr                             & 4      & 19   \\
Cu$_{2}$NiSiCr                       & 5     & 27   \\
Cu$_{2}$Ni$_{2}$Si$_{2}$Cr$_{2}$     & 8     & 2404    \\
Cu$_{3}$NiSiCr                       & 6     & 100    \\
Cu$_{3}$Ni$_{2}$Si$_{2}$Cr           & 8     & 1571    \\
Cu$_{4}$NiSiCr$_{2}$                 & 8     & 8   \\
Cu$_{4}$NiSi$_{2}$Cr                 & 8     & 16   \\
Cu$_{4}$Ni$_{2}$SiCr                 & 8     & 8  \\
Cu$_{5}$NiSiCr                       & 8     & 4    \\
\hline
\end{tabular}
\label{table:data1}
\end{table}

\section{Methodology\label{sec:method}}

In this work, we apply the computational scheme and workflow we have successfully applied  for the prediction of binary phases in copper alloys \cite{carral2023stability}. In short, our workflow consists of (a) quantum-mechanical simulations on a large pool of relevant structures, (b) the calculation of the convex hull based on the formation enthalpies of these structures, and the use of data from quantum-mechanical simulations for (c) actively generating machine learned interatomic potentials (MLIPs) in order to (d) perform subsequent classical simulations of the predicted stable phases and reveal their properties. Here, this scheme is used to predict and propose novel potentially stable quaternary phases that are very close to the convex hulls. Details of the applied scheme can be found in Ref.\,\citenum{carral2023stability}. In the following, we only mention the most important aspects of this framework, which is applied here to assist the prediction of novel ternary Ni-Si-Cr  phases from the pool of phases that make up Cu-Ni-Si-Cr alloys.

In order to sample the configuration space of Cu-Ni-Si-Cr phases with a vast number of stoichiometries and calculate the relevant energies, we turn to a sampling scheme generating the corresponding structures. For this,  a certain class of MLIPs, the Moment Tensor Potentials (MTPs) \cite{mtp1} are developed on the basis of an on-the-fly active learning algorithm and density functional theory (DFT) based simulations \cite{GUBAEV2019148}. We refer to this scheme in the following as AL-MTP. The input of AL-MTP consists of: a set of structure candidates to be geometrically optimized, an MTP functional form $ E = E(\Theta, x) $, outputs from DFT simulations, and two learning thresholds $ \gamma_\text{tsh} $ and $ \Gamma_\text{tsh} $, with $ \Gamma_\text{tsh} > \gamma_\text{tsh} > 1 $. In the MTP functional form $E$, $ \Theta $ represents the MTP parameters, while the function is determined by the level $\text{lev}_\text{max}$. The latter  denotes the level of complexity or approximation utilized in constructing the basis functions and affects the accuracy, computational cost, and transferability of the MTPs. For the generated quaternary structures, the convex hulls for the various  Ni-Si-Cr phases in a Cu matrix are calculated with respect to their formation enthalpies and stoichiometries. The quaternary phases predicted here using the AL-MTP scheme will be discussed in Section \ref{sec:result} and investigated though classical Molecular Dynamics (MD) simulations on the basis of the MTPs generated in order to assess selected properties of the discovered quaternary phases. The generation of candidate quaternary structures and the exploration of the relevant phase space of stoichiometries was enabled by the use of the ENUMLIB library \cite{enumlib2}. The latter could generate 4159 prototype quaternary structures as listed in Table \ref{table:data1}. Each structure corresponds to a distinct arrangement of atoms within the unit cell, the smallest repeating unit in the crystal lattice that defines the material's symmetry and composition. The number of atoms in the unit cell ("cell size" in the table), is crucial in determining the properties of the generated phases. These samples were used as a starting library for the AL-MTP scheme, supplemented by an additional fcc structure reported in the Open Quantum Materials Database (OQMD) \cite{oqmd2,kirklin2015open}.

\begin{table}[h!]
\centering
\caption{Ground-state atomic energies ($E_{i}$) in eV/atom for the alloying elements $i$=\{Cu, Ni, Si, Cr\} in different crystal symmetries (symm), with $N_i$ atoms in the unit cell, and the magnetic state (magn).}
\setlength{\tabcolsep}{6pt} 
\begin{tabular}{|p{0.8cm}|p{1.2cm}|p{1.2cm}|p{2cm}|p{1.2cm}|}
\hline
{$i$} & symm & {$N_i$} & {$E_{i}$ (eV/atom)} & magn \\
\hline
Cu    & fcc        & 4       & -3.710  & NM  \\
Ni    & fcc        & 4       & -5.566  & FM  \\
Si    & fcc        & 8       & -5.424  & NM  \\
Cr    & bcc        & 2       & -9.631  & AFM \\
\hline
\end{tabular}
\label{tab:totalenergies}
\end{table}

We begin this work with DFT simulations of the unit cells of Cu, Ni, Si, and Cr in order to find the respective ground-state atomic energies, which correspond to the thermodynamically stable unit cell structures. These were performed with the DFT implementation in the code  VASP \cite{vasp1996, paw1999}. We have used the Perdew-Burke-Erzernhof (PBE) \cite{perdew1996generalized}  generalized gradient approximation for the description of the exchange-correlation functional and the projector augmented wave method (PAW) \cite{blochl1994projector} for the pseudopotentials of Cu, Ni, Si, and Cr. We have considered the magnetic behavior of the alloying elements, including non-magnetic (NM), ferromagnetic (FM), and antiferromagnetic (AFM) states by performing spin-polarized simulations. The DFT calculations were performed with a plane-wave cutoff energy of 650 eV  for the  basis set description and a k-point spacing of approximately 0.157  \r{A}$^{-1}$ for proper Brillouin zone sampling. We employed a smearing method to handle electronic partial occupancies in metallic systems and incorporated spin polarization to accurately model magnetic properties. Additionally, we utilized stress tensor calculations to optimize the cell volume. The convergence threshold for the forces on the ions was set to 0.01 eV/\AA~ and for the energy to 10$^{-4}$ eV. Periodic boundary conditions were applied in all three directions.

In order to determine the optimal cell volumes, we performed a scan of the lattice constant vs. energy space for all the alloying elements, which allowed the cell volume to change, thereby effectively optimizing the lattice parameters. The ground-state atomic energies as calculated using DFT are provided in Table \ref{tab:totalenergies}. The formation enthalpy ($\Delta H_{f}$) of the quaternary phases depends on the ground-state atomic energies through \cite{leicester1951germain}
\begin{equation}
\Delta H_{f} = \frac{1}{N}\left[E_{tot} - \sum^{4}_{i}\frac{n_i}{N_{i}}\cdot E_{i}^{}\right],
\label{eq:hess}
\end{equation}
where $N$ and  $E_{tot}$ are  the total number of atoms and the  energy for each quaternary structure, respectively, $n_i$ is the number of atoms in the unit cell for species $i$ in the structure, $N_i$ is the number of atoms of the unit cell of species $i$ in the respective one-component system. AL-MTP is used to predict the total energy, $E_{tot}$, while $E_{i}^{}$ is calculated via volume relaxation for the distinct unit cell symmetries of each element using DFT. The calculated formation enthalpies are used to construct the convex hull of the quaternary structures.

\begin{table}
  \centering
   \caption{Fitting  errors during the AL-MTP process, evaluated through the mean average error and the root-mean-squared-error, 'MAE' and 'RMSE', respectively (both in meV/atom), the
    training set size ('training'), i.e. number of configurations, for the fcc-derivative sets for different lev$_\text{max}$ 16, 18 and 20, respectively. 
   }
  \begin{tabular}{|p{2.cm}|p{2cm}|p{2.cm}|p{2.cm}|}
    \hline
    {lev$_\text{max}$} & {training} & {MAE} & {RMSE ($\sigma$)}       \\
    \hline
       16            & 1858      & 9.265    & 12.178        \\
       18            & 2238      & 8.617    & 11.784        \\
       20            & 2731      & 7.782    & 10.471        \\
  \hline
  \end{tabular}
  \label{tab:mtperror}
\end{table}

\begin{figure}
\centering
\includegraphics[width=\columnwidth]{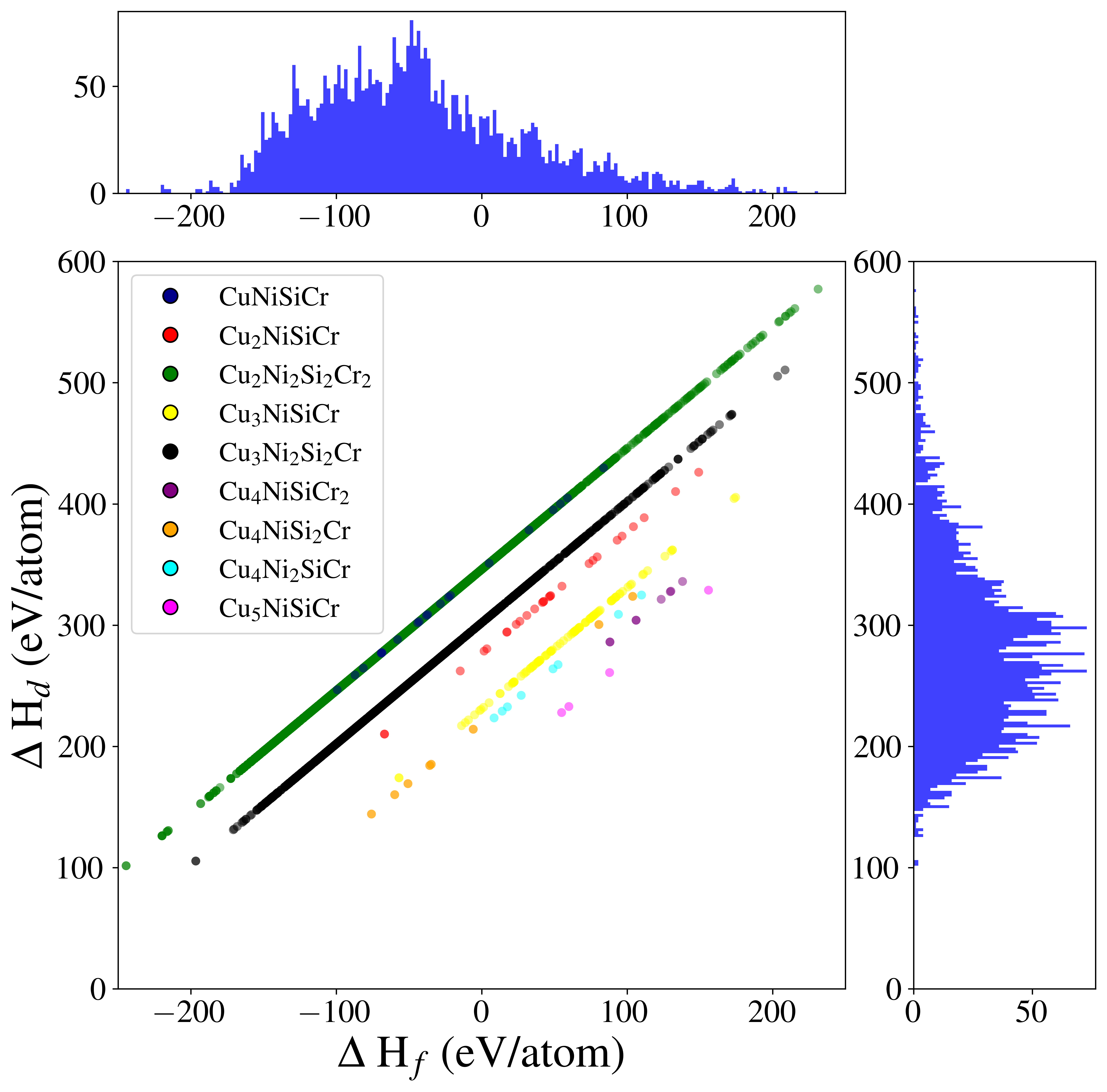}
\caption{A scatter plot of the relationship between the formation enthalpy ($\Delta H_f$) and the hull distance  ($\Delta H_d$) for selected groups in the CuNiSiCr alloys, as denoted in the legend. On the top and right of the graph, the respective probability distributions of $\Delta H_{f}$ and $\Delta H_{d}$, respectively, are shown. 
}
\label{fig:deltaHvsdeltaHd}
\end{figure}

The convex hull (or its projection) provides information on the relative stability of the materials. An important quantity that assesses the thermodynamic phase stability is the distance to the hull $\Delta H_{d}$, also known as the decomposition energy. The energy distance of a compound to the convex hull offers insights into its absolute stability or metastability and the feasibility of its synthesis. This measure also provides valuable information about the uncertainties associated with these factors, further enhancing our understanding of the material's overall characteristics. Metastable materials near the convex hull may undergo transformation into more stable phases during prolonged aging \cite{cheng2014evaluation}. The rate at which these transformations occur can be influenced by kinetic barriers associated with atomic rearrangements. In fact, the probability to synthesize a material decreases rapidly with $\Delta H_{d}$, which not only indicates that a material is unstable, metastable or stable, but also quantifies the stability itself, providing insights into the uncertainty of stability assessment and synthesis plausibility. In characterizing the formation enthalpy across various stoichiometries, the Grand Canonical Linear Programming (GCLP) method has been applied \cite{GCLP1,GCLP2}. GCLP effectively tackles the complexity of metal hydride thermodynamics by utilizing linear algebra to solve free energy minimization problems. This approach is applicable to structures representing a high-dimensional phase space allowing the visualization of formation energy data, thus offering valuable insights into the intricate thermodynamic landscape of high-entropy alloys. More specifically, the reference values of the formation enthalpy for the unique stoichiometries presented in Table \ref{table:data1} are calculated using the GCLP module available in the OQMD repository.

The AL-MTP scheme has the potential to predict novel (i.e. not yet synthesized) (meta)stable structures, which are in the current context further modeled using classical Molecular Dynamics (MD) simulations implemented in the code LAMMPS with the LAMMPS-MLIP interface \cite{LAMMPS, mlip2}. MD additionally assesses the stability of the structures, as well as their thermal and elastic properties and dynamical behavior or time evolution. The initial atomic coordinates for the MD simulations were obtained by replicating a stable relaxed unit cell predicted through the AL-MTP scheme implemented in this work. Periodic boundary conditions were used in all three directions of space. The protocol followed in the MD calculations included two relaxation steps while keeping separately the pressure and temperature of the system constant, a longer relaxation, followed by the production run from which the properties of the structures are evaluated and discussed in the following. Specifically, MD simulations were performed in the anisotropic isothermal-isobaric ensemble (NPT). Different target temperature values in the range [100:1300]\,K were used. An initial equilibration step of 10\,ps was performed in the canonical ensemble (NVT), where the temperature $T$ was controlled using a Berendsen thermostat with a time constant of 0.1\,ps. This was followed by a second equilibration step of 20\,ps in the NPT ensemble, where the pressure $p = 1$\,bar was maintained using a Nose-Hoover barostat with a time constant of 1\,ps, and the temperature $T$ was controlled by a Nose-Hoover thermostat with a time constant of 0.1\,ps. Then, production runs lasting 500\,ps were conducted in the NPT ensemble under the same conditions as the second equilibration step. A timestep of 0.5\,fs was used for all three steps. The simulation time was proven sufficient for the solid structures investigated here. This was also confirmed by the well-resolved radial distribution functions (RDFs) provided in the end, suggesting a high signal-to-noise ratio. From the production runs, several observables were calculated, including the RDFs and the density $\rho = M / V$, where $M$ is the total mass of the system and $V$ its volume. The mean square displacement $\text{MSD} = \langle | \mathbf{x} (t) - \mathbf{x_0} |^2 \rangle$ of the Si impurity was also computed. Finally, measurements of the bulk $B$ and shear $\mu$ moduli, elastic constants $c_{ij}$, and Poisson's ratio $\nu$ were performed at $T=0$\,K. LAMMPS input scrips are provided in the GitHub repository together with our AL-MTP potential; \cite{simon_gravelle_2024_13341067}.

\begin{figure}
\centering
\includegraphics[width=\columnwidth]{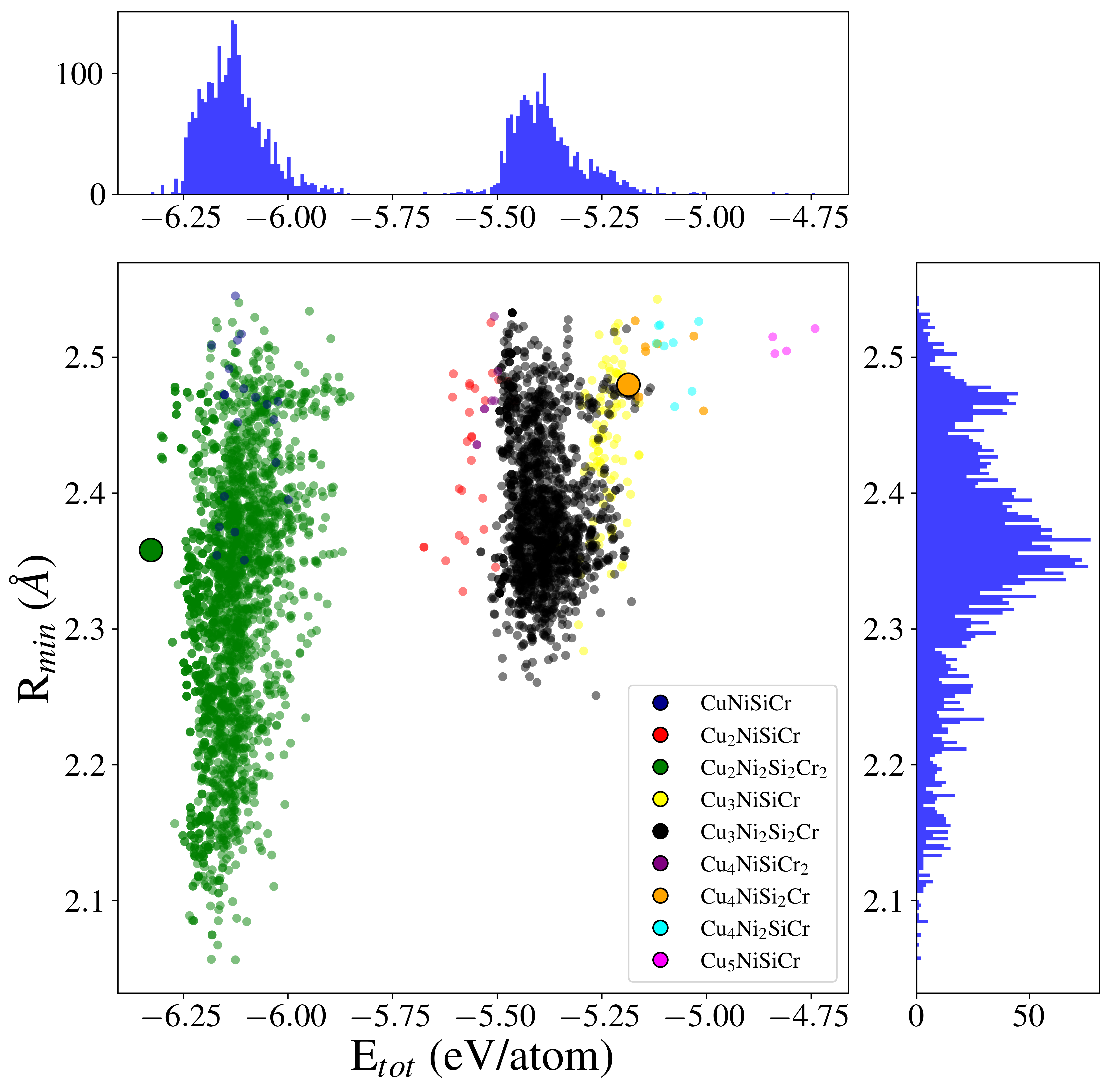}
\caption{A scatter plot depicting the connection between the total energy (E$_{tot}$) and the unit cell distance R$_{min}$ from AL-MTP in CuNiSiCr alloys with different stoichiometries. Each point signifies a group, color-coded according to the stoichiometry, with the chemical compositions described in the legend.  Two relevant structures identified by AL-MTP are highlighted (large circles) with their respective colors according to the stoichiometry. On the top and right of the graph, the respective probability distributions of the R$_{min}$ and E$_{tot}$ are shown.}
\label{fig:d_ene}
\end{figure}

\section{Results and Discussion \label{sec:result}}

The methodology outlined above allows a detailed study of the Cu-Ni-Si-Cr alloy, where fcc-rich regions in a Cu matrix \cite{cheng2014evaluation} were found along with the well known Cr-Si and Ni-Si binary precipitates \cite{ma15134521}. A first assessment of the stability of quaternary phases in a copper matrix relies on the evaluation of their calculated formation enthalpy from Eq.\ref{eq:hess}. In this assessment, we rely on the $4\sigma$ rule, with $\sigma$ being the root-mean-square error (RMSE) of the energy per atom measured on the AL-MTP training set. This allows the refinement of all modelled structures.  We implement a post-relaxation strategy, targeting structures with formation energies below $4\sigma$ from the convex hull. 
Note, that is it computationally demanding to calculate the  \( E_{\text{tot}} \) in Eq.\ref{eq:hess} for all 4159 configurations using DFT. In order to follow a more efficient scheme, we first relax these structures using AL-MTP to  obtain the total energy for each structure. We then select the most promising structures based on the 4-sigma rule and post-relax them using DFT to ensure accuracy. By comparing the \( E_{\text{tot}} \) values from both DFT and MTP, we demonstrate that MTP can achieve quantum mechanical (QM) accuracy while being significantly faster. This comparison is meaningful because we evaluate the same energy (\( E_{\text{tot}} \)) using two different methods, providing similar results. Accordingly, this comparison ensures the reliability and efficiency of the MTPs as an alternative to DFT calculations of  formation enthalpies and  convex hull construction.

In Table \ref{tab:mtperror}, we present learning information on the generated AL-MTP potentials for different values of $ \text{lev}_{\text{max}} $: 16, 18, and 20. Specifically, the training set size refers to the number of pre-selected data points by the AL-MTP approach, demonstrating the potential of this method. For instance, for an MTP of level 16, we could predict the total energy of all the candidates by computing 1,858 DFT static calculations, compared to 4,159 DFT relaxations required if these had been relaxed. The table also provides the mean absolute error (MAE) calculated during the fitting process. This approach addresses challenges linked to outliers and computational artifacts, ensuring that the selected alloys maintain thermodynamic feasibility. In order to screen the energy landscape of the generated dataset on the quaternary copper alloys, we relax all structures following the combined AL-MTP approach and first discuss the respective enthalpies depicted in Fig.\ref{fig:deltaHvsdeltaHd}. In this figure, the formation enthalpy of all  stoichiometries implied in Table \ref{table:data1} is correlated to the  respective hull distance. Each point in the figure represents a distinct stoichiometry group, color-coded accordingly. At first, this graph  provides a visual insight into the energetic stability and structural variations within the respective configurations. In the same figure, the histograms for the formation enthalpy and the hull distances computed are also depicted. 

\begin{figure}
\centering
\includegraphics[width=\columnwidth]{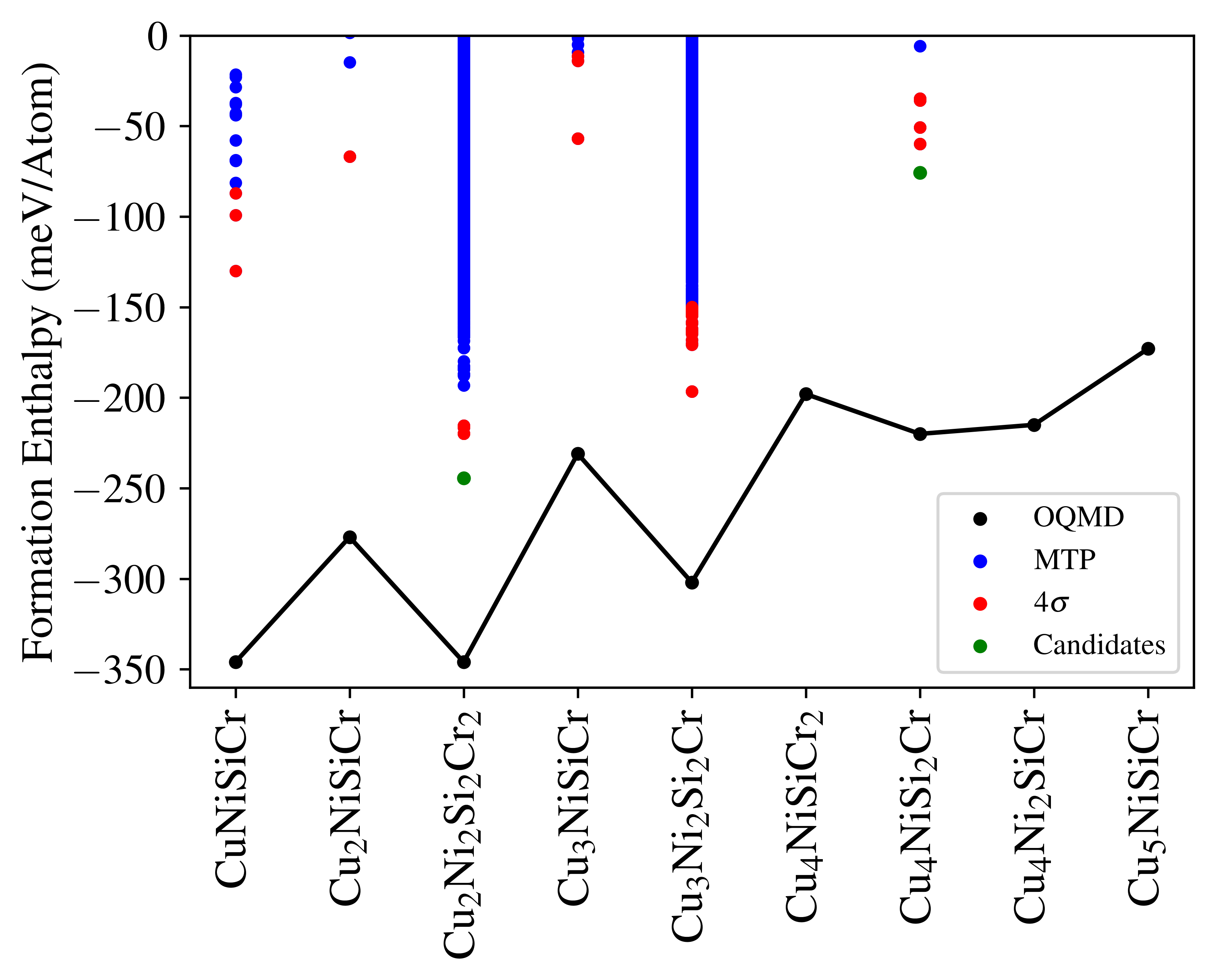}
\caption{The  convex hull projected with respect to the different CuNiSiCr phases. The respective formation enthalpy is shown for the structures in the OQMD database, the MTP relaxed, and the 4$\sigma$ post-relaxed with DFT, as labelled in the graph  (see text for details).
}
\label{fig:chulls}
\end{figure}

\begin{table*}
\centering
\caption{Lattice vectors a, b, c (\(\text{\AA}\)), angles $\alpha$, $\beta$, $\gamma$ (\textdegree), volume (V) (\(\text{\AA}\)$^3$), lattice symmetry (LS), and space-group (SG) of the discovered phases. The formation enthalpy (meV/atom) is given for both the AL-MTP relaxation (AL-MTP) and the post-relaxation with DFT (DFT). 
}
\begin{tabular}{|p{2.1cm}|p{1.2cm}|p{1.2cm}|p{1.2cm}|p{1.3cm}|p{1.3cm}|p{1.3cm}|p{1.3cm}|p{1.3cm}|p{1.3cm}|p{1.3cm}|p{1.3cm}|}
\hline
{phase}& {a} & {b/a} & {c/a} & {$\alpha$} & {$\beta$} & {$\gamma$} & {V} & {LS} & {SG}  & {AL-MTP}  & {DFT}  \\
\hline 
tet-CuSiNiCr  &  2.918 & 3.698 & 1.000 & 90 &  90 &  90  & 91.892 & TET & \textit{P4/nmm} & -244.46  & -247.61\\ 
Cu$_{4}$NiSi$_{2}$Cr   &  5.212 & 1.000 &  1.000 &  95.436 & 123.187 &  56.813  & 94.843 & BCT  & \textit{I4/mmm} & -75.82 & -77.55 \\
\hline
\end{tabular}
\label{tab:prototypeslattice}
\end{table*}

Inspection of Fig.\ref{fig:deltaHvsdeltaHd} reveals a linear correlation of the formation enthalpy of each structure with its hull distance. At the same time, the various stoichiometries are well-separated in the $\Delta H_d$-$\Delta H_f$ two-dimensional space. Interestingly, as the stoichiometry index of the copper atom in the unit cell of the quaternary structure increases, the corresponding correlation lines in the figure shift towards smaller hull distances and higher formation enthalpies. This trend is clearly evident by focusing on the different colors in the figure, as long as only the copper stoichiometry changes. Once one of the impurities is also increased, this shift moves back closer to the respective lower Cu concentration. In order to support this, one can compare the correlation lines of the CuNiSiCr to Cu$_2$NiSiCr and the latter to Cu$_2$Ni$_2$Si$_2$Cr$_2$. Evidently, the concentration of copper in the alloy increases the formation enthalpy of the structure, thus moving it to the metastable and unstable region. On the other hand, the simultaneous increase of the impurity concentration seems to stabilize the alloy. Although the number of candidates with higher stoichiometries decreases, still these trends persist. The observed trends are also mirrored in the hull distance. Specifically, as the concentration of copper and sequentially that of the impurities increases, the distance of the average point of the respective stoichiometry to the convex hull decreases, as well as the formation enthalpy. A visual insight into the energetic stability and structural variations within the various configurations is provided in  Fig.\ref{fig:d_ene}. Interestingly, the structures with the same stoichiometry are mapped through well-defined clusters in the space of the minimum distance and the total energy. This maps the structural similarity of the respective materials. Specifically, two are the more rich clusters, namely those related with the Cu$_{2}$Ni$_{2}$Si$_{2}$Cr$_{2}$ and the Cu$_{3}$Ni$_{2}$Si$_{2}$Cr, which are the ones with 2 nickel and silicon atoms in the unit cell. In addition, most of the modelled quaternary phases have a  unit cell distance in the range [2.3:2.4] \AA.

\begin{figure}
\centering
\includegraphics[width=\linewidth]{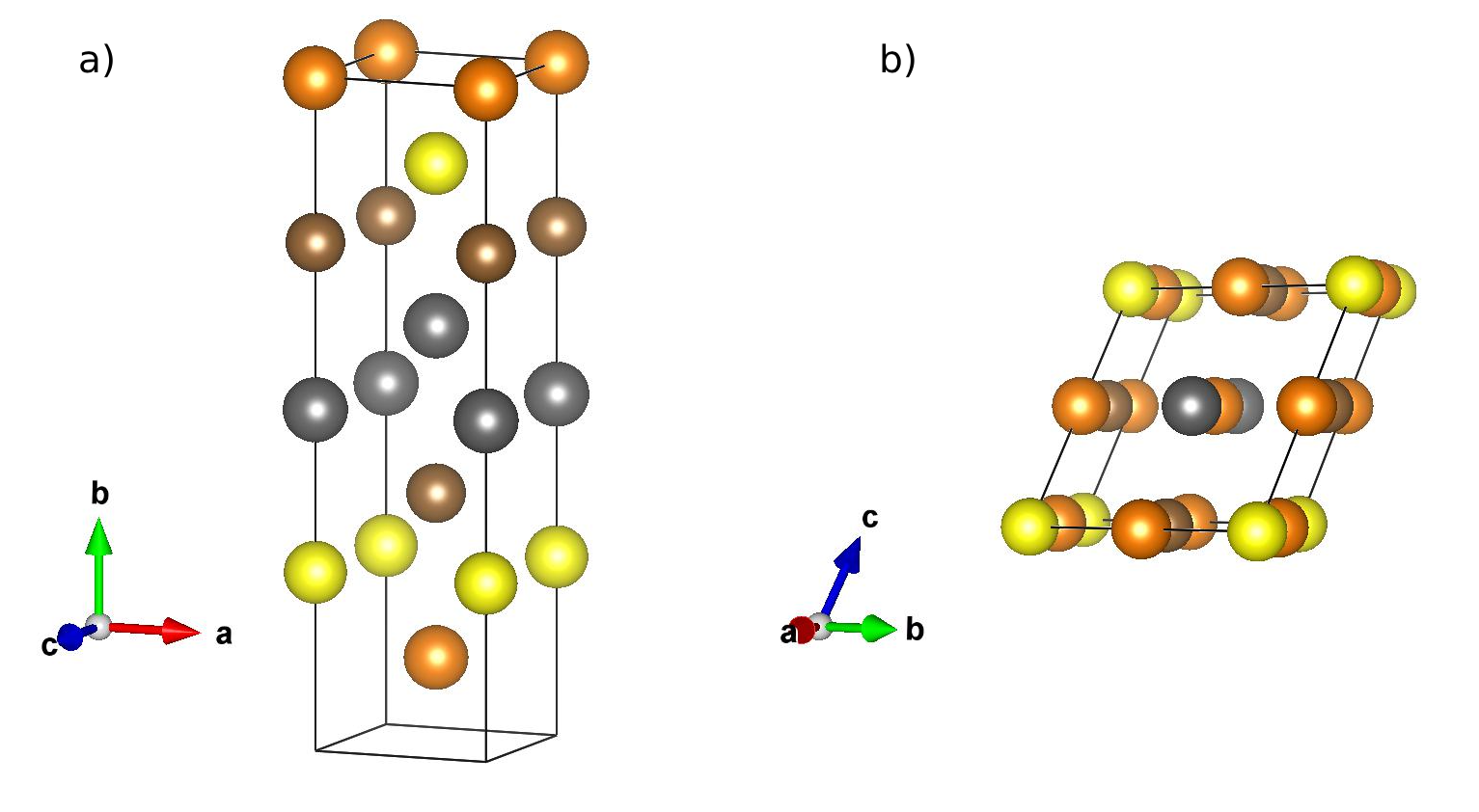}
\caption{The unit cells of the predicted (a) tet-CuSiNiCr and (b) Cu$_{4}$NiSi$_{2}$Cr. The Cu, Ni, Si, and Cr atoms are depicted in orange, yellow, brown and gray, respectively.}
\label{fig:structures}
\end{figure}

\subsection{Prediction of novel quaternary structures\label{sec:novel}}

The convex hull as projected on the different quaternary Cu-Ni-Si-Cr phases is provided in Fig.\ref{fig:chulls}. This projection was performed based on the higher dimensional convex hulls for the formation enthalpy and the stoichiometries of the Ni, Si, and Cr impurities.  The formation enthalpy values shown refer to those based on the OQMD database, the MTP-calculated values, and the values derived from the $4\sigma$ rule mentioned above. Specifically, we post-relaxed the prototype CuNiSiCr discovered in the OQMD library and the $4\sigma$ prototypes highlighted in red, focusing on the one exhibiting the lowest formation enthalpy within the entire dataset and the most stable structures with a composition of Cu$_4$NiSi$_2$Cr. This stoichiometry closely aligns with the experimental observations of the fcc (Ni, Cr, Si)-rich phase reported previously \cite{cheng2014evaluation}. The post-relaxation with DFT of all structures provides higher formation enthalpy values than the OQMD formation enthalpies (projected convex hull). However, we could identify two new stoichiometries in the DFT post-relaxed structures that are linked to formation enthalpies closer to two of the minima in the OQMD values,  as highlighted in Fig. \ref{fig:d_ene}. These are the 4$\sigma$ post-relaxed values closest to the DFT post-relaxation structures for the stoichiometries Cu$_{2}$Ni$_{2}$Si$_{2}$Cr$_{2}$  and Cu$_{4}$NiSi$_{2}$Cr.  In Fig.\ref{fig:chulls}, the green points correspond to the two relevant structures identified by the AL-MTP scheme, referred to as `candidates'. The formation enthalpies of the two candidates after the MTP relaxation are closer to the convex hull projection in this figure than before relaxation. Note that Fig. \ref{fig:chulls} visualizes the position of each of the 4159 relaxed structures with respect to the convex hulls. It demonstrates the metastability of all these supporting the experimental findings \cite{cheng2014evaluation}, the choice of the two most relevant metastable configurations, as well as the consistency of the MTP approach in this choice for further DFT validation. Accordingly, we move on our analysis by focusing on the metastability of the structures close to the theoretically stable OQMD convex hull (black curve), which can also be of practical interest under certain conditions. With the convex hull, we also highlight the 4$\sigma$ post-relaxed structures closest to the hull (red points) and the two most relevant metastable ones (green points) with respect to their stoichiometry. These structures were re-evaluated using DFT to verify their (meta)stability and remain metastable. 

\begin{figure}
\centering
\includegraphics[width=\linewidth]{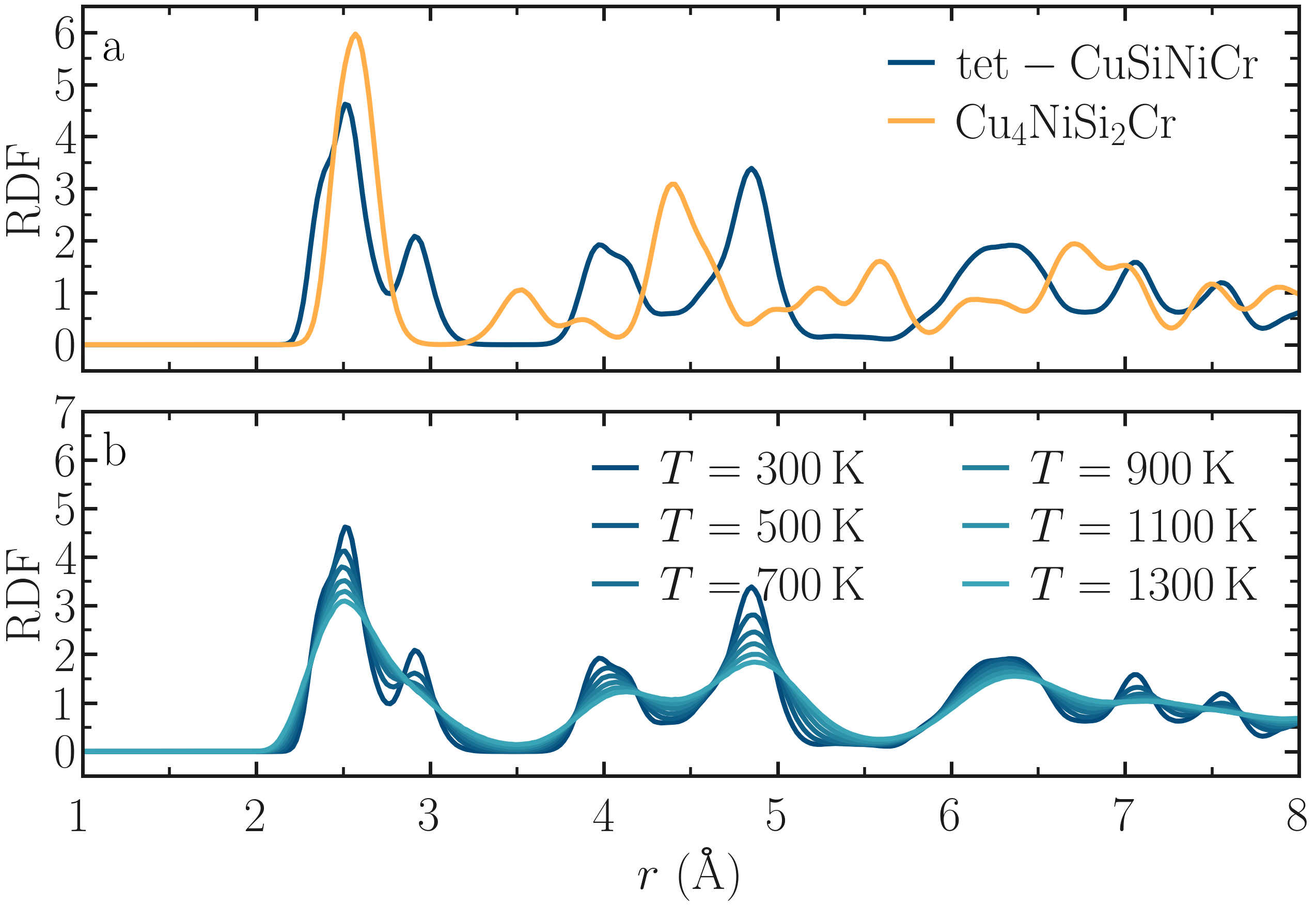}
\caption{a) RDFs as calculated for both quaternary structures tet-CuSiNiCr and Cu$_{4}$NiSi$_{2}$Cr using molecular dynamics for a temperature $T = 300$\,K. b) RDFs for temperatures corresponding to $T=300$, 500, 700, 900, 1100, and $1300$\,K for the tet-CuSiNiCr structure.
}
\label{fig:rdf-md}
\end{figure}

Accordingly, we were able to identify two novel quaternary phases, the structural details of which are depicted in Table \ref{tab:prototypeslattice}, and the respective unit cells are visualized in Fig.\ref{fig:structures}. These two novel phases are labelled as tet-CuSiNiCr and Cu$_{4}$NiSi$_{2}$Cr, respectively. Note that we have labelled the discovered Cu$_{2}$Ni$_{2}$Si$_{2}$Cr$_{2}$ as tet-CuSiNiCr, since the primitive Bravais lattice exhibits a tetragonal (tet) symmetry. As a derivative superstructure, its lattice vectors are multiples of a cubic (fcc) parent lattice before post-relaxation, and its atomic basis vectors correspond to lattice points in the parent lattice. The choice of the two structures is based on the following argument: First, the point with the smallest $\Delta H_d$ represents the most metastable structure within our generated dataset. Second, the proposed Cu${4}$NiSi${2}$Cr is the closest candidate for that stoichiometry, which closely resembles the experimentally observed composition \cite{cheng2014evaluation}. A greater proximity to the convex hull suggests increased metastability and implies that these structures are expected to remain stable for a longer time under specific kinetic conditions during aging. If a metastable phase can be synthesized and is retained kinetically \cite{kineticmeta}, the respective alloy can be of a high practical interest under certain conditions. Note, that it is difficult to compute the convex hulls of highly multi-component materials. Typically, the more components in a material, the more metastable and less stable these are.

\begin{figure}
\centering
\includegraphics[width=\linewidth]{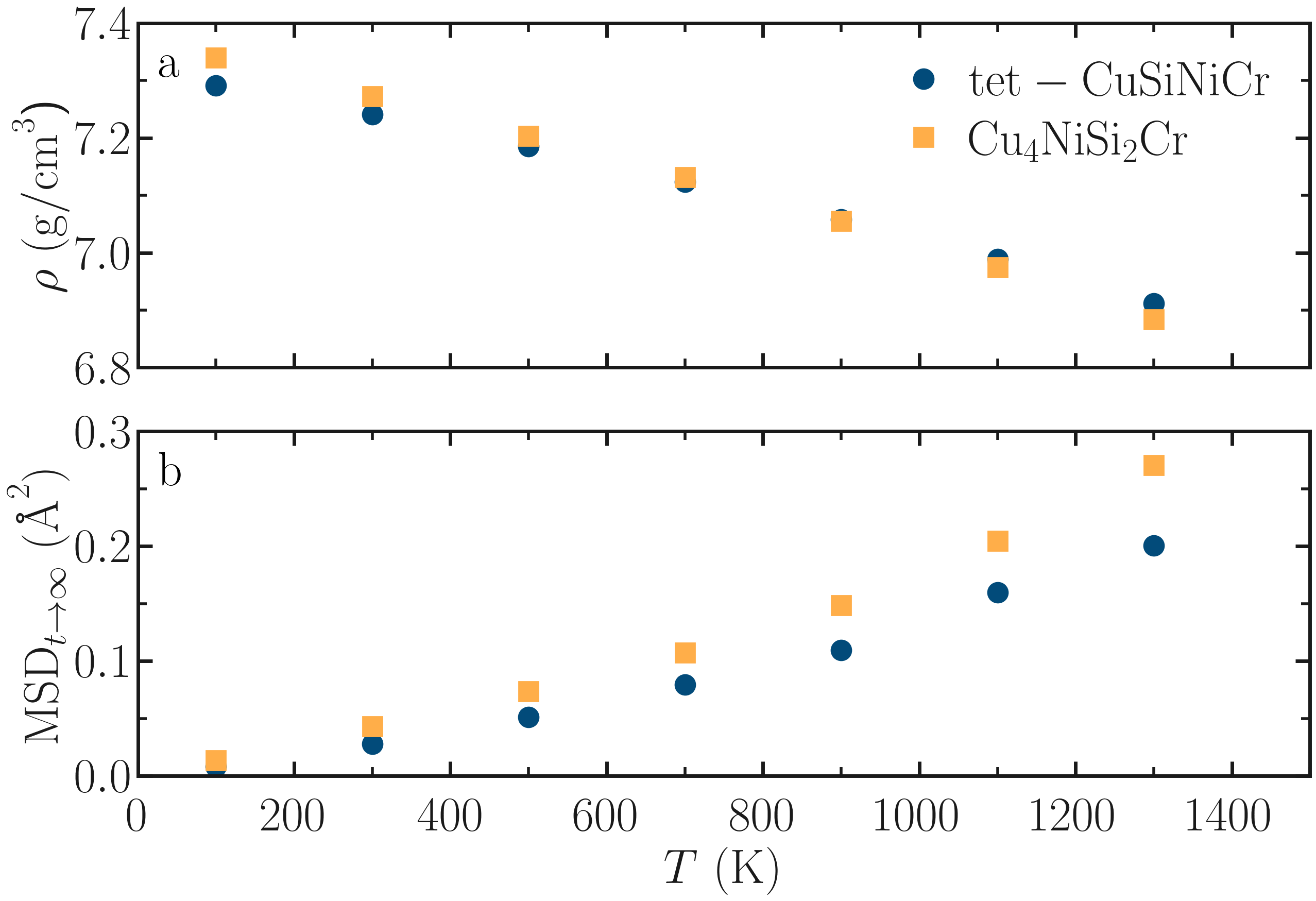}
\caption{a: Density $\rho$ as calculated for both quaternary structures tet-CuSiNiCr and Cu$_{4}$NiSi$_{2}$Cr using molecular dynamics as a function of the temperature. b: Mean square displacement (MSD) of the Si atoms in the long time limit ($t \to \infty$).
}
\label{fig:density-msd-md}
\end{figure}

\begin{table*}
\caption{The LAMMPS-MLIP calculated bulk modulus $B$, elastic constants $c_{ij}$, shear modulus $\mu$, and Poisson's ratio $\nu$ for the two discovered materials. All quantities are given in GPa, except for the Poisson's ratio that is unitless.} 
\centering
\begin{tabular}{|p{2.5cm}|p{1.5cm}|p{1.6cm}|p{1.6cm}|p{1.6cm}|p{1.6cm}|p{1.6cm}|p{1.5cm}|p{1.4cm}|p{1.4cm}|}
\hline
& $B$ & $c_{11}$  & $c_{22}$  & $c_{33}$  & $c_{12}$  & $c_{13}$  & $c_{23}$ & $\mu$  & $\nu$ \\
\hline 
tet-CuSiNiCr & 263.07 & 351.94 & 270.79 & 351.95 & 215.43 & 265.62 & 215.43 & 146.86 & 0.416 \\
Cu$_{4}$NiSi$_{2}$Cr & 166.13 & 262.31 & 277.95 & 293.39 & 126.29 & 112.20 & 92.25 & 62.6 & 0.284 \\
\hline
\end{tabular}
\label{tab:elastic-properties}
\end{table*}

\subsection{Structural dynamics and elastic behavior}

The developed machine learned potentials are further implemented at the classical  scale of atomistic MD simulations. The generated MTPs applied on the supercells of the predicted structures are used to calculate the thermodynamic and mechanical properties of the quaternary materials. This scheme carries with it the quantum-mechanical accuracy from the DFT simulations, and is thus better than the empirical force-fields. As a proof of concept, in the following the results are obtained by MD simulations using the MTPs developed above during the active learning scheme. The MD simulations are performed for both predicted quaternary phases in Cu alloys, with the computational parameters discussed in the Methodology. 

The RDFs were computed at different temperatures for both phases and are summarized in Fig. \ref{fig:rdf-md}. An inspection of the top panel of this figure reveals peaks of different heights and at different positions for the tet-CuSiNiCr and Cu$_{4}$NiSi$_{2}$Cr phases. This again confirms the different (bonding) environment in these structures. In order to unveil the influence of the temperature, we choose the tet-CuSiNiCr structure and simulate its dynamics for different temperatures in the range $[300:1300]$\,K as implied in Fig. \ref{fig:rdf-md}b. An increasing temperature $T$ leads to a broadening of the peaks. This intuitively points to larger atomic deviations from the crystalline positions. Specifically, at temperatures greater than or equal to $\approx 1500$\,K, our results (not shown) indicate a melting of both quaternary structures. In this regime, though, the simulations were numerically unstable, suggesting that the developed MTPs are more accurate for modeling solid phases and need to be adjusted to cover also the liquid phases.

In order to support the gradual melting of the predicted quaternary phases and follow this process, we monitor the changes in the density of the structures, as well as the mean square displacements of the atoms in these. The respective density $\rho$ of both solid structures is provided as a function of the temperature in Fig.\ref{fig:density-msd-md}a. Our results show that $\rho$ decreases with increasing temperature, from $\rho \approx 7.1$\,g/cm$^3$ for $T = 100$\,K, to $\rho \approx 6.9$\,g/cm$^3$ for $T = 1300$\,K. The trends observed are almost identical for both structures apart from the start and end temperatures. This indicated that, though, the composition in these phases is different, apparently also the atomic densities, the total density behaves in a very similar manner. To evaluate differences at the atomic level, we discuss the mean square displacement (MSD) of the Si atoms in Fig.\ref{fig:density-msd-md}b. The Si atoms are selected to present the displacement of the smaller, thus more mobile atoms, in these structures. Accordingly, the MSD results for the Si atoms represent the upper bound to the corresponding values for the other atomic species in the predicted phases. Note that the results are shown in the limit of long time scales. Inspection of the MSD results show very similar trends for both quaternary phases, as the two curves increase in the same way with increasing temperatures. However, resolving the atomic scales unravels the inherent differences in the two alloys, with the tet-CuSiNiCr showing less mobility at the atomic level than the Cu$_{4}$NiSi$_{2}$Cr. This trend can be traced back to the higher content of Si atoms in the latter case.

Finally, in order to also provide insight into the mechanical properties of the predicted quaternary phases, we provide in Table\,\ref{tab:elastic-properties}, their elastic moduli  as calculated from the MD simulations. To this end, the bulk  $B = (c_{11}+2 c_{12})/3$ and shear $\mu = c_{44}$ moduli, Poisson's ratio $\nu$, and selected elastic constants ($c_{11}$, $c_{22}$, $c_{33}$, $c_{12}$, $c_{13}$, $c_{23}$) are summarized in this table. For all moduli, the values for tet-CuSiNiCr are much higher than those for Cu$_{4}$NiSi$_{2}$Cr. In fact, in average, these are two times larger. These trends imply a softer Cu$_{4}$NiSi$_{2}$Cr material than tet-CuSiNiCr, which can be more easily deformed along all modes and directions. This trend goes hand-in-hand with the trends in Fig.\ref{fig:density-msd-md}: a higher mobility results in a more deformable material.

\section{Conclusions \label{sec:concl}}

Using a combination of quantum-mechanical and classical computer simulations together with active machine learning, we have predicted two novel stable quaternary phases relevant to copper alloys, including nickel, silicon, and chromium as impurities. During this process, we were able to not only predict new phases, but also develop interaction potentials, which we further use to model the physical properties of the predicted quaternary phases. We have analyzed their structural, dynamic, and elastic properties and identified insightful trends in view of the practical use of the respective alloys. It is important to note that the (meta)stability predicted through our computational work does not directly guarantee successful experimental synthesis. In practice, materials may encounter kinetic barriers that prevent their formation, and experimental challenges could limit the realization of these structures. However, with the AL-MTP approach used here, it is possible to further refine the materials design by investigating the potential transformation of the (Ni, Cr, Si)-rich clusters into the $\delta$-Ni$_{2}$Si phase over an extended aging period, incorporating both thermodynamic and kinetic factors alongside experimental validation. Ultimately, this approach is highly efficient, combining quantum-mechanical accuracy, active learning, and atomistic dynamics. It can be further optimized to provide deeper insights into stability and transformation pathways, and to better integrate with experimental studies. Our findings underline the computational efficiency of the combined simulation and Machine Learning approach in materials prediction and larger-scale calculations of their properties with a quantum-mechanical accuracy. Specifically, we show the accuracy of these tools in the case of four-component materials, which can be applied to other types of structures and extended to multi-component materials to provide valuable insights into the structure-property relations in materials in view of novel technological applications.

\section{Data availability statement\label{sec:state}}

The data that support the findings of this study are available from the corresponding author upon reasonable request. The generated machine-learned potentials, as well as the LAMMPS inputs, are available in the GitHub repository; \cite{simon_gravelle_2024_13341067}.

\section*{Conflict of interest}

The authors have no conflicts to disclose.

\begin{acknowledgments}
The authors are grateful to C. Holm and the Institute for Computational Physics at the University of Stuttgart in Germany for support. A.D.C. and M.F. acknowledge financial support from the EXC 2075 SimTech Cluster of the University of Stuttgart. S.G. acknowledges funding from the European Union's Horizon 2020 research and innovation programme under the Marie Skłodowska-Curie grant agreement N$^\circ\;101065060$.
\end{acknowledgments}

\bibliography{bibliography}{}
\bibliographystyle{plain}

\end{document}